# ICT and Employment in India: A Sectoral Level Analysis

(Dr. Pawan Kumar)

Assistant Professor, Ramjas College
University of Delhi, Delhi-110007
E.mail: pawankumar@ramjas.du.ac.in

## (Abstract)

How technology affects growth or employment has long been debated. With a little hiatus, the debate revived once again to evaluate the impact of Information and Communication Technology (as a form of new technology) on productivity and employment. ICT perceived as GPT (General Purpose Technology) like steam engine or electricity in the past, ushered the world into a new *techno-economic paradigm.* It is hard to imagine an economic activity without using it, directly or indirectly. Eventually, ICT investment as percentage of non ICT investment (known as ICT intensity) increased significantly over the years in industries across all sectors.

What impact ICT intensity has on employment is an important question taken up in the paper. To find the answer, on the basis of ICT intensity, industries belonging to the organized sector are categorized into ICT producing (ICTPS), ICT using (ICTUS) and non ICT using (NICTUS) sectors, with their further division into manufacturing and services sector. Empirically, it is found that only the ICTPS witnessed high employment elasticity with increased ICT intensity since 2000, something found true in both of its segments manufacturing and services sectors, and in both periods, Period I (200-05) and Period II (2005-05). In contrast, in ICTUS, the employment elasticity declined with increased ICT intensity. The trend found true in both of its constituents. Finally, within NICTUS, in both periods, in both sectors, it is the same nature of employment elasticity with increased ICT intensity.

In conclusion, for the country as a whole, employment elasticity has increased, driven by services sector, with increased ICT intensity. So, it can be ascertained that new technology in the form of ICT has resulted in positive employment impact in services sector, but not in the secondary sectors.

**Introduction**

The relationship between technology and employment is complex and controversial. In the last decade, Information and Communications Technology (ICT), in the form of new technology, has emerged in a big way the world over including India. It is believed that after the technological waves of *steam engine* in the 19th century or and *electricity* in the 20th century, ICT is the only technology being regarded as general purpose technology (GPT). With the growth rate of over 25 percent since 2006, ICT producing sector has grown in size and in employment generation, particularly for skilled workers.

Till the early 1980s, countries particularly belonging to the OECD group were apprehensive while using ICTs. There was an apprehension that the impact of the relatively higher investment made in ICT did not percolate into higher growths of employment and productivity a notion later known as *productivity paradox*- that computers are seen everywhere except in productivity statistics-a concept introduced and popularized by Solow (Soete, 1987). The perception about the impact of ICT on employment changed over time.

In the 1990s empirical studies mostly based on the case studies of the U.S. and some countries of the EU concluded and established the positive and significant impact of ICT on productivity and employment. This, however, gave rise to yet another debate, why productivity (TFP) or employment growth rose faster in the U.S. than in the EU. The debate got somewhat resolved towards the mid-2000s, with a general consensus emerged that it is the *ICT producing industries* (particularly the services sector) that triggered productivity and employment growth in the U.S. post-1995; something found missing in the EU (Vivarelli, 2011). In other words, compared to the EU countries did not seem to have exploited the productivity enhancing potential of the ICT-producing industries to the extent possible. The results marked a major departure from results of the earlier studies that doubted the potential of the ICT led productivity growth of the services sectors (Gordon, 2000).

Three channels are identified through which ICT influences growth; surge in ICT investment, strong productivity effects from ICT-producing industries and spillover impacts in the ICT using sectors of the economy (Vivarelli and Pianta, 2000). Though the ICT spillovers are typically difficult to measure at industry level but at firm-level they are found to be present.

In India, the success story of ICT growth began gradually since the early 2000s, it accelerated after 2005. It has become a major source of foreign exchange earning through foreign investment (FDI and FIIs) and export of ITES (IT enabled services). According to NASSCOM (2012), in 2008-09, the sector grew by 14 percent to reach $ 71.7 billion in aggregate revenue (including hardware). Of this, the software and services segment accounted for a major chunk ($ 59.6 billion). In the same year, the total ICT revenue reached 5.8 percent of GDP compared to 1.2 percent in 1997-98. Despite some slowdown, the Indian IT sector has successfully weathered the global financial crisis of 2008-09.

Not much research has been done on association between ICT, productivity and employment as far as the developing countries (including India) are concerned (Freeman and Soete, 1987 and Vivarelli, 2011). Often, success stories of the newly industrialized countries (NICs) like, South Korea, Philippines, Indonesia, Malaysia, etc. are cited as examples when it comes to evaluate or measure the contribution of ICT in growth or employment (OECD, 2010). In India, reports of NASSCOM and Planning Commission, or some other anecdotal studies though exist, but they deal within the general framework of socio-economic implications of ICT such as ICT impact on environment, on women empowerment, on rural development, on skill, and so on. In the paper, an attempt is made to evaluate the rise of ICT intensity across industries and the corresponding employment growth in India.

Towards this, industries, belonging to the organized sector in India are divided into three groups, ICT producing sectors (ICTPS), ICT using sectors (ICTUS) and non-ICT using sectors (NICTUS), with their additional sub-division into manufacturing and services sectors respectively.

The paper is structured as follows. It starts with the definition of ICT and ICT intensity, followed with data source and research methodology. Thereafter, an analysis of the behavior of output growth, employment growth, employment elasticity and ICT intensity is made for all the three groups of industries.

1. **Review of Literature**

After a lot of debate and discussion, ICT was defined as: *The production (goods and services) of a candidate industry that must primarily be intended to fulfill (or enable) the function of information processing and communication by electronic means, including transmission and display (OECD, 2010).* Given the pervasive nature of ICT, it is considered as GPT or as a form of new technology.

Theoretically, there is no direct (or well defined) way to know the impact of technology employment. The contradictory nature of results on how technology affects employment arise due to different assumptions made about the output growth rate, demand, different level of aggregation (sector, industry or firm) and the way it's indirect impacts is treated, i.e. whether these effects are included or not in the whole analysis (Bhalla, 1997 and Kumar, 2005). More so, it also depends how technology is defined, disembodied or embodied. The former is proxied by MFP, and the latter is embodied in factors of production, labour or capital. For example, in neo classical economics, technical change is measured by MFP.

Empirically, the views on how technology affects employment are broadly categorized as optimistic and pessimistic. In the former based on the *compensation theory,* it is believed that any technical change with the implicit presence of various *compensation mechanisms* always results in positive employment impact, at least, in the long run (Vivarelli, 2011). It is evident from the results in many OECD countries, particularly the U.S during 1960-2000 that technological change is accompanied by increased employment growth, with enhanced MFP (Multi-factor productivity) growth (Stiroh, 2002 and Vivarelli, 2011). The pessimistic view, on the other hand, though, in principle, believe in the working of *compensation theory* but rule out the possibility of the complete counter balance of the labour saving (or negative employment) impact of technical change. The great economist Wassily Leontief worried that the pace of modern technological change is so rapid that many workers, unable to adjust, will simply become obsolete, like horses after the rise of the automobile (Rogoff Kenneth, WEF, 2012).

In case of the ICT, there is a growing apprehension that it has weakened the positive correlation between growth, productivity and employment something that has been one of the main characteristics of the post-World War II period known as 'golden age' (Rifkin, 1995). Many countries, developed and developing, have been experiencing structural unemployment (or this weakening relationship) originating from ICT (Vivarelli and Pianta, 2000).

In last two decades, ICT emerged in a big way the world over including India. Accepted as a general purpose technology (GPT), ICT led the world economy ushered into a new *techno-economic paradigm.* However, its influence on productivity or employment varies across firms, industries and countries. The pioneering hypothesis, in this regard, was made by Freeman, Clark and Soete (1982) that ICT is good not only for productivity but for employment as well. *Adding further, it concluded that structural unemployment in E.U. during the 1970s was not due to ICT but due to socio-institutional rigidities.*

OECD (2010) defined many ways through which ICT affects employment growth, classified as direct and indirect. Harrison, et. al. (2006) investigated the effect of ICT on employment growth in a number of OECD counties; **they concluded that during 1998-00, ICT resulted in direct and positive employment impact, and it also resulted in positive indirect employment impact following compensation mechanism.** However, as far as the indirect employment impact of ICT is concerned it is found that it will result into a loss of nearly 5-10 million jobs annually the world over (Rogoff Kenneth, WEF, 2012). Surprisingly, labour market is able to absorb these losses in some other sectors.

Further, in the empirical studies, on the basis of the way technology is used a distinction is also made between product and process innovations. For instance, in the latter, when the new product is used as capital in the production process of other goods or services, then technology is regarded as a job killer. In the former, new or improved product is used as final product in consumption, and then certainly it will be a job creator, if not met with demand deficiency (Schmidt, 1983; Pissarides and Vallanti, 2003). *When measuring the employment impact of ICT, ICT, as new technology, is used as a product innovation in the ICT producing sectors and as a process innovation in the ICT using sectors (OECD, 2010). In the former, its employment impact is immediate, direct and positive, i.e., more is the output growth more is the employment assuming no demand deficiency; and in the latter it is indirect, positive (or negative) and often arises in long run. The net employment impact is therefore positive or negative depends on the relative sizes of these two effects.* Empirical studies conducted on many OECD countries, particularly the U.S. and the E.U. concluded that in the former the ICT producing sector is stronger both in terms of MFP and employment growth.

In India, contribution of ICT in GDP and employment is well documented (NASSCOM, 2012). Nevertheless, there are apprehensions that it has resulted in negative employment in sectors using it. In other words, it is believed that problem of unemployment got further accentuated with rise in ICT use.

Results at the aggregate level may at times be misleading; it can obscure the reality at the disaggregated level (i.e. at industry or at the firm level). The study is therefore conducted at industrial level as well. Towards this, total number of industries is classified into the ICT-producing (ICTPS) and the ICT using industries (ICTUS) and non ICT using industries; categorized further into manufacturing and services units. The ICTPS includes producers of IT hardware, communication equipment, telecommunications and computer services (including software). The distinction between ICTUS and NICTUS was made on the basis of the level of ICT intensity. Practically, no industries can however be classified as Non-ICT industry, every industry uses some bit of ICT directly or indirectly. Theoretically, ***an industry is defined as non ICT if the ICT intensity is lesser than one third of the national average.***

When measuring the impact of ICT on employment, the concept of employment elasticity is used to find the extent of labour demand for every unit of good produced. Employment elasticity (EE) is defined as the ratio of employment and output growth rates. It measures the employment content of the additional unit of output produced. In other words, EE is defined by the formula,

$$EE=(dL/L)/dY/Y)$$

Where, L stands for employment while Y denotes value of output. The numerator is interpreted as the percentage change of employment, while the denominator refers to the percentage change in output. The EE shows the degree of responsiveness of employment to a percentage change in GDP. It is used to understand the labour market dynamics, and is therefore used in labour policy formulation.

## 2. Classifications of Industries into ICT Producing, ICT Using and Non ICT Using Industries

The idea of classification of industries into ICT Producing, ICT Using and Non ICT Using Industries is derived from Stiroh (2002). He used the relative size of the contribution of these groups in order to provide a plausible explanation for the productivity difference in the U.S. and the Europe. Further, Van ark, et. al. (2002) extended the same definition to 16 OECD countries and 49 industries. The hypothesis underlying the geographical extension of the original classification introduced by Stiroh (2002) is that in each industry ICT would face the same pattern of adoption across countries, considering the U.S. as the optimal pattern. With the original 118 industries in total, 13 were under ICT producing industries, 49 under ICT using and the remaining 66 under non ICT using category.

In India, because of the data constraints, only 89 industries are considered; of which 9 are under the ICT producing sector, 39 in the ICT using sector and the remaining 41 under non ICT using category. Like the original classification, three categories are sub-classified into in the secondary and service sector; this is follows: 6 and 3; 19 and 20; and 31 and 10 respectively in ICTPS, ICTUS and NICTUS.

### 3. Output Growth, Employment Growth, Employment Elasticity (EE) and ICT Intensity in India: a Group Level Analysis

In this section an attempt is made to analyze the behavior of output growth rate, employment growth rate, employment elasticity (EE) and ICT intensity of the groups identified as ICTPS, ICTUS and NICTUS. The total period (2000-10) is divided into two sub-periods Period I (2000-2005) and Period II (2005-10). The analysis is made first at the aggregated and then at the disaggregated level i.e. at groups level. *Employment and Output growth rates are given as compound annual growth rates (CAGR).*

As shown in Table 1, the Indian economy at the aggregated level registered an impressive output growth rate of 12 percent since 2000, with 10 percent and 13 percent in Period I and Period II respectively. Similarly, the employment growth rate is found to be 3 percent at the aggregated level with respectively 2.5 and 4 percent in Period I and Period II. Eventually, EE increased from 0.25 to 0.29. What led this to happen, and to what extent is this attributed to the increase to new technology (or ICT intensity), which has gone up remarkably from 2 percent in Period I to 8 percent in Period II. In order to capture comprehensive picture, the study is extended to the di-aggregated level.

**3.1 ICT Producing Group (ICTPS)**: Among all, the ICTPS group posted the highest growth rates of employment and output in both periods. As shown in Table1, the former, accelerated from 9.5 percent in Period I to13.4 percent in Period II, and the latter also remained at double digit level 30 percent and 17 percent respectively in Period I and Period II. It led EE increased from 0.31 in Period I to 0.75 in Period II. This means, if output has doubled it resulted in labour demand by 31 percent in Period I but by 75 percent in Period II. The group, ICTPS, also recorded the maximum increase in ICT intensity from 2.2 percent in Period I to 11.6 percent in Period II.

In India, in both sub-groups of ICTPS, i.e., in secondary and services sectors, rising ICT intensity is followed by increased employment growth. For instance, the former recorded output growth rate 16 percent in Period II up from 9 percent in Period I; and employment growth rate 5 percent from 2 percent. Consequently, EE rose from 0.2 to 0.31. This sub-group recorded increased ICT intensity from 3 percent in Period I to 6 percent in Period II. Further, as shown in Table 2, all six industries except industry group 323 (Manufacture of TV and radio receivers and recording or reproducing apparatus) posted accelerated growth rates of output and employment in Period II. The EE within the secondary sub-group, however, shows a mixed trend.

The *services sub-group* of ICTPS has witnessed the highest growth rates in output and employment and ICT intensity. For instance, as shown in Table 3, output and employment growth rate are estimated to be 40 percent and 35 percent respectively in Period I though they declined to 19 percent and 22 percent in Period II, but these are still highest among all groups and sub-groups. Though EE declined marginally in Period II but it remained around one. Further, as evident in Table 3, the trend of impressive growth rate is recorded by all its constituent industries, 642 (Telecommunications), 722 and 723 (Software Consultancy and Data Processing).

The results outlined above in case of the ICTPS group hence indicate that the positive relationship between ICT and employment. This in a sector in which new technology is used as a product, ICT product innovation; thus satisfying the **compensation mechanism through new products** *that new technology always brings new employment opportunities in the sector producing it if there is no demand deficiency (Vivarelli, 2011).*

### 3.2 ICT Using Sectors (ICTUS)

The ICTUS group consists of 39 industries in total, 19 in secondary and the rest 20 in services sector. The group has ICT intensity more than non ICT group but less than the ICT producing group, and can therefore be termed as an *intermediate group*. In the group, a comparative analysis of growth rates of employment and output show that it recorded these rates more than in the NICTUS but less than the ICTPS. EE is recorded to have declined in Period II. The important question is what led this to happen?

At the dis-aggregated level, the secondary sub-group of ICTUS witnessed more or less the same trend as the group at a whole be it the output or the employment growth rate, and so the EE. EE declined sharply from 0.36 to 0.23 as evident in Table 1. ICT intensity is found to have increased from 2 percent in Period I to 3.3 percent in Period II. Further, the analysis at the industrial level within the sub-group shows the same pattern of growths in output and employment as the results at the aggregated level of the sub-group. In other words, be it publishing (221), re-production of recorded media (223) or transport equipment (359), all have recorded lower EE with increased ICT intensity.

Similarly, the services component of the ICTUS group also displayed the same trend as the secondary sub-group, increasing output growth rate accompanied by decreasing employment growth rate, and thus lower EE (from 0.36 in Period I to .26 in Period II). Further, as evident from Table 5, results at the industrial level show a mixed trend in EE, whereas ICT has gone up across all industries.

As discussed above, *the ICTUS use ICT as **process innovation** in the form of Compute Aided Design/Computer Aided Machines; and most empirical studies show, industries using ICT as process innovation experience negative employment growth rate. Most countries of OECD countries experienced the same results as far as the relationship between ICT and employment is concerned (Pianta, 2005 and Brynjolfson, 2004).*

### 3.3 Non ICT Using Sector (NICTUS)

The NICTUS group, which by definition has the least ICT intensity, includes 40 industries in total (30 from secondary and 10 from services). As shown in Table 1, both employment and output rates accelerated but EE declined marginally from 0.5 in Period I to 0.4 in Period II. ICT intensity has increased, though from a lower base.

Further, as evident in Table 01 the secondary sub-group of NICTUS witnessed employment and output growth rates changing from 4.14 percent and 7 percent respectively in Period I to 4.8 percent and 8 percent respectively in Period II. Given this, EE declined from 0.60 to 0.56. ICT intensity is recorded to have gone up. But, as evident in Table 6, industries in the secondary sector show a mixed trend of EE. Similarly, the services sub-group has also recorded a marginal decline in EE in Period II. Contrary to this, industries in the sub-group have portrayed a mixed trend of increase in EE. Interestingly, all industries have experienced increased ICT intensity in Period II as shown in Table 07.

From the above description, it is clear that ICT intensity has some relationship with employment, but at this stage it is difficult to comment on the nature of this relationship, i.e., what is the net employment impact of ICT intensity at the sectoral level, group level and at the aggregated level. In the next section, an attempt is made to asses this.

**Conclusion**

Since 2000, development and diffusion of ICT is on the rise the world over including India. ICT, accepted as GPT (General Purpose Technology) has led the Indian economy into a *new techno-economic paradigm.* The study conducted empirically in India gives many important results on the relationship between new technology and employment. Starting with the ICTPS group, with ICT intensity employment growth has also gone up since 2000; something found true for both secondary and services sub-groups also. Though, empirically it is beyond the scope of this paper, the extent of employment gain or loss due to ICT intensity, the theory of *compensation mechanism* help us to explain the results found. **First**, as mentioned above, the *compensation mechanism through new products* such as semi-conductor, computer hardware and telecommunication devices always result in positive employment impact. **Second,** in India, if ICT is used as *product innovation,* it leads to positive employment impact. **Third,** *compensation via decline in prices* of ICT product is also found strong in India. In other words, as per Moore's Law, the power of semi-conductor gets doubled every 20 months while its price halves, give a major boost in the demand ICT and related product and, therefore, more employment. **Fourth,** *compensation via increase in investment* also helped in increased employment growth. For instance, ICT sector has recorded comparatively high investment growth, contributed significantly by FDI inflow. For instance, over 600 Multinational companies (MNCs) are known to be sourcing their product development and engineering services from their centers in India (GOI, 2007-08).

Like ICTPS, employment growth in the ICTUS group in India has gained positively from new technology, but the impact is found significant only in Period II. A significant share in total output and employment comes from the ICT using services sector. In the group, ICT is used as *process innovations*, which has negative employment impact through replacement of highly labour intensive electromechanical work with increasingly integrated components produced by automation produced in other manufacturing component. Conversely, in the services sub-group has witnessed a positive employment of ICT intensity. It may be due to all the other *compensations mechanisms* working in its favor.

Finally, in the NICTUS group, in secondary and services sectors, in India, impact of ICT on employment may not be very strong, because of low ICT intensity base.

**Table1: Employment Growth, Output Growth, Employment Elasticity and the ICT Intensity (%) in the Secondary Sector in India during 2000-10**

| Industries | Years | E | Y | EE | ICT/Non ICT (%) |
|---|---|---|---|---|---|
| **All Industries** | 2000-10 | **2.98** | **11.65** | **0.26** | 4.73 |
|  | 2000-05 | 2.56 | 10.22 | 0.25 | 1.99 |
|  | 2005-10 | 3.76 | 12.94 | 0.29 | 8.14 |
| **ICT Producing Sectors (ICTPS)** | 2000-10 | 11.15 | 24.48 | 0.46 | 6.38 |
|  | 2000-05 | 9.4 | 30.65 | 0.31 | 2.22 |
|  | 2005-10 | 13.33 | 17.76 | 0.75 | 11.58 |
| ICT Producing Secondary Sectors (ICTPMS) | 2000-10 | 2.67 | 9.46 | 0.28 | 3.73 |
|  | 2000-05 | 1.88 | 9.34 | 0.20 | 2.59 |
|  | 2005-10 | 4.90 | 15.61 | 0.31 | 5.53 |
| ICT Producing Services Sectors (ICTPSS) | 2000-10 | 27.86 | 28.83 | 0.97 | 6.55 |
|  | 2000-05 | 40.56 | 34.66 | 1.17 | 1.97 |
|  | 2005-10 | 18.98 | 21.54 | 0.88 | 12.29 |
| **ICT Using Sectors (ICTUS)** | 2000-10 | 3.55 | 11.91 | 0.30 | 4.05 |
|  | 2000-05 | 3.34 | 12.01 | 0.28 | 3.28 |
|  | 2005-10 | 3.05 | 13.79 | 0.22 | 5.01 |
| ICT Using Secondary Sectors (ICTUMS) | 2000-10 | 3.04 | 10.88 | 0.28 | 2.49 |
|  | 2000-05 | 3.01 | 8.36 | 0.36 | 1.87 |
|  | 2005-10 | 3.27 | 14.02 | 0.23 | 3.27 |
| ICT Using Services Sectors (ICTUSS) | 2000-10 | 3.58 | 12.85 | 0.28 | 5.15 |
|  | 2000-05 | 3.73 | 10.37 | 0.36 | 3.9 |
|  | 2005-10 | 3.29 | 12.69 | 0.26 | 6.72 |
| **Non ICT Using Sectors (NICTUS)** | 2000-10 | 4.02 | 9.5 | 0.42 | 0.35 |
|  | 2000-05 | 3.91 | 8.38 | 0.47 | 0.20 |
|  | 2005-10 | 4.22 | 10.89 | 0.38 | 0.66 |
| Non ICT Using Secondary Sectors (NICTMS) | 2000-10 | 4.29 | 7.55 | 0.62 | 0.39 |
|  | 2000-05 | 4.14 | 6.88 | 0.60 | 0.18 |
|  | 2005-10 | 4.77 | 8.39 | 0.56 | 0.65 |
| Non ICT Using Services Sectors (NICTUSS) | 2000-10 | 4.5 | 15.75 | 0.28 | 0.45 |
|  | 2000-05 | 4.43 | 15.08 | 0.29 | 0.24 |
|  | 2005-10 | 4.6 | 16.49 | 0.27 | 0.70 |
| **Total Secondary Sectors** | 2000-10 | **4.31** | **9.96** | **0.43** | 2.9 |
|  | 2000-05 | 4.01 | 8.19 | 0.49 | 1.87 |
|  | 2005-10 | 4.98 | 12.09 | 0.41 | 4.35 |
| **Total Services Sectors** | 2000-10 | **6.36** | **20.89** | **0.30** | 4.71 |
|  | 2000-05 | 6.09 | 20.02 | 0.30 | 1.8 |
|  | 2005-10 | 5.94 | 16.90 | 0.35 | 8.19 |

Source: Own Computation using the Proves data source compiled by CMIE

**Table 2: Employment Growth, Output Growth, Employment Elasticity and the ICT Intensity (%) ithe ICT Producing Secondary Sector**

| NIC 2004 | | Year | E | Y | EE | ICT/Non ICT (%) |
|---|---|---|---|---|---|---|
| 300 | Office Accounting and Machinery of Computers | 2000-10 | 7.19 | 9.36 | 0.77 | 8.60 |
| | | 2000-05 | 6.38 | 9.04 | 0.71 | 5.05 |
| | | 2005-10 | 8.22 | 9.77 | 0.84 | 13.05 |
| 313 | Insulated Wires, Cables and Computers | 2000-10 | 7.29 | 11.05 | 0.65 | 1.37 |
| | | 2000-05 | 6.28 | 10.66 | 0.58 | 0.63 |
| | | 2005-10 | 9.53 | 12.77 | 0.75 | 2.29 |
| 321 | Electronic Valves, Tubes and Other Electronic Components | 2000-10 | 5.85 | 7.91 | 0.74 | 4.72 |
| | | 2000-05 | 5.85 | 5.83 | 1.00 | 2.37 |
| | | 2005-10 | 5.86 | 10.75 | 0.54 | 7.66 |
| 322 | TV and Radio Transmitters and other Communications Apparatuses | 2000-10 | 2.82 | 8.21 | 0.34 | 5.57 |
| | | 2000-05 | 2.17 | 3.97 | 0.54 | 3.50 |
| | | 2005-10 | 4.31 | 13.50 | 0.32 | 8.17 |
| 323 | TV and radio receivers and recording or reproducing apparatus | 2000-10 | 0.48 | 15.13 | 0.03 | 3.64 |
| | | 2000-05 | 1.58 | 17.29 | 0.09 | 3.42 |
| | | 2005-10 | 0.88 | 12.42 | 0.07 | 3.92 |
| 331 | Medical and Surgical Instruments and Orthopedics Appliances | 2000-10 | 6.81 | 12.37 | 0.55 | 2.63 |
| | | 2000-05 | 5.04 | 10.48 | 0.48 | 3.18 |
| | | 2005-10 | 8.52 | 15.74 | 0.54 | 1.94 |
| Aggregate | | 2000-10 | 2.67 | 9.46 | 0.28 | 3.73 |
| | | 2000-05 | 1.88 | 9.34 | 0.20 | 2.59 |
| | | 2005-10 | 4.90 | 15.61 | 0.31 | 5.53 |

Source: Own Computation using the Proves data source compiled by CMIE

**Table 3: Employment Growth, Output Growth, Employment Elasticity and the ICT Intensity (%) in the ICT Producing Services Sector**

| NIC 2004 | Industry | | E | Y | EE | ICT/Non ICT (%) |
|---|---|---|---|---|---|---|
| 642 | Telecommunications | 2000-10 | 27.47 | 32.11 | 0.86 | 3.17 |
| | | 2000-05 | 41.58 | 42.84 | 0.97 | 1.46 |
| | | 2005-10 | 14.82 | 18.70 | 0.79 | 5.31 |
| 722+723 | Software Consultancy and Data Processing | 2000-10 | 29.71 | 26.63 | 1.12 | 34.74 |
| | | 2000-05 | 39.63 | 27.27 | 1.45 | 6.59 |
| | | 2005-10 | 17.31 | 25.82 | 0.67 | 69.93 |
| Aggregate | | 2000-10 | 27.86 | 28.83 | 0.97 | 6.55 |
| | | 2000-05 | 40.56 | 34.66 | 1.17 | 1.97 |
| | | 2005-10 | 18.98 | 21.54 | 0.88 | 12.29 |

Source: Own Computation using the Proves data source compiled by CMIE

**Table 4: Employment Growth, Output Growth, Employment Elasticity and the ICT Intensity (%) in the ICT Using Secondary Sector during 2000-10**

| NIC 2004 | Industries | Years | E | Y | EE | ICT/Non ICT (%) |
|---|---|---|---|---|---|---|
| 181 | Wearing apparel and accessories expect fur apparel | 2000-10 | 11.71 | 16.81 | 0.70 | 5.05 |
| | | 2000-05 | 10.38 | 12.49 | 0.83 | 2.33 |
| | | 2005-10 | 13.37 | 22.21 | 0.60 | 8.45 |
| 221 | Publishing | 2000-10 | 5.93 | 12.18 | 0.49 | 3.01 |
| | | 2000-05 | 9.83 | 12.97 | 0.76 | 0.83 |
| | | 2005-10 | 1.05 | 11.19 | 0.09 | 5.73 |
| 223 | Reproduction of recorded Media | 2000-10 | 8.59 | 11.62 | 0.74 | 2.41 |
| | | 2000-05 | 6.26 | 6.62 | 0.95 | 0.90 |
| | | 2005-10 | 11.50 | 17.88 | 0.64 | 4.29 |
| 291 | machinery for the production and use of mechanical power, except aircraft vehicle and cycle engine | 2000-10 | 1.21 | 13.29 | 0.09 | 5.16 |
| | | 2000-05 | 1.21 | 10.44 | 0.12 | 2.03 |
| | | 2005-10 | 4.23 | 16.85 | 0.25 | 9.07 |
| 292 | other general purpose machinery | 2000-10 | 1.11 | 5.18 | 0.21 | 2.17 |
| | | 2000-05 | 1.71 | 2.27 | 0.75 | 0.87 |
| | | 2005-10 | 0.37 | 8.82 | 0.04 | 3.80 |
| 293 | domestic appliance | 2000-10 | 1.06 | 4.46 | 0.24 | 2.40 |
| | | 2000-05 | 0.83 | 4.12 | 0.20 | 0.81 |
| | | 2005-10 | 3.41 | 4.88 | 0.70 | 4.39 |
| 311 | electric motors, generators and transformers | 2000-10 | 3.08 | 11.71 | 0.26 | 1.50 |
| | | 2000-05 | 2.30 | 6.57 | 0.35 | 0.95 |
| | | 2005-10 | 4.05 | 18.14 | 0.22 | 2.19 |
| 312 | electricity distribution and control apparatus | 2000-10 | 4.63 | 20.62 | 0.22 | 2.46 |
| | | 2000-05 | 4.73 | 18.05 | 0.26 | 1.59 |
| | | 2005-10 | 5.02 | 23.82 | 0.21 | 3.54 |
| 314 | accumulators, primary cells and primary batteries | 2000-10 | 3.02 | 11.93 | 0.25 | 2.00 |
| | | 2000-05 | 1.59 | 4.70 | 0.34 | 0.98 |
| | | 2005-10 | 4.80 | 20.97 | 0.23 | 3.27 |
| 315 | lighting equipment and electric lamp | 2000-10 | 4.46 | 10.02 | 0.45 | 1.73 |
| | | 2000-05 | 1.57 | 2.70 | 0.58 | 1.19 |
| | | 2005-10 | 8.08 | 19.16 | 0.42 | 2.41 |
| 319 | other electric equipment | 2000-10 | 0.95 | 4.86 | 0.19 | 1.46 |
| | | 2000-05 | 0.56 | 1.62 | 0.34 | 0.62 |
| | | 2005-10 | 2.83 | 8.91 | 0.32 | 2.52 |
| 321 | electronic valve, tubes and other electronic component | 2000-10 | 2.08 | 6.14 | 0.34 | 1.26 |
| | | 2000-05 | 0.75 | 4.48 | 0.17 | 0.51 |
| | | 2005-10 | 3.75 | 8.23 | 0.46 | 2.20 |
| 351 | Building and repairing of ships and boats | 2000-10 | 3.31 | 14.62 | 0.23 | 0.35 |
| | | 2000-05 | 1.35 | 3.15 | 0.43 | 0.10 |
| | | 2005-10 | 9.14 | 28.96 | 0.32 | 0.66 |
| 359 | transport equipments | 2000-10 | 2.84 | 10.34 | 0.27 | 1.90 |
| | | 2000-05 | 4.49 | 11.88 | 0.38 | 1.27 |
| | | 2005-10 | 0.79 | 8.42 | 0.09 | 2.68 |
| 361 | Manufacture of furniture | 2000-10 | 16.37 | 17.40 | 0.94 | 2.10 |
| | | 2000-05 | 27.04 | 24.68 | 1.10 | 1.65 |
| | | 2005-10 | 3.04 | 8.29 | 0.37 | 2.67 |
| 369 | Manufacturing n.e.c. | 2000-10 | 5.73 | 22.86 | 0.25 | 3.95 |
| | | 2000-05 | 6.60 | 24.30 | 0.27 | 1.85 |
| | | 2005-10 | 4.66 | 19.81 | 0.23 | 6.57 |
| 401 | Production, transmission and distribution of electricity | 2000-10 | 8.06 | 4.00 | 2.01 | 0.06 |
| | | 2000-05 | 11.09 | 4.99 | 2.22 | 0.05 |
| | | 2005-10 | 4.26 | 2.76 | 1.54 | 0.08 |
| 452 | Building of complete constructions, civil engineering | 2000-10 | 23.27 | 25.40 | 0.91 | 2.49 |
| | | 2000-05 | 22.85 | 19.82 | 1.15 | 2.02 |
| | | 2005-10 | 24.05 | 32.39 | 0.74 | 3.09 |
| 453 | Building installation | 2000-10 | 9.06 | 16.11 | 0.56 | 1.33 |
| | | 2000-05 | 7.97 | 19.11 | 0.47 | 0.40 |
| | | 2005-10 | 10.42 | 24.86 | 0.41 | 2.49 |
| Total | | 2000-10 | 3.04 | 10.88 | 0.28 | 2.49 |
| | | 2000-05 | 3.01 | 8.36 | 0.36 | 1.87 |
| | | 2005-10 | 3.27 | 14.02 | 0.23 | 3.27 |

Source: Own Computation using the Proves data source compiled by CMIE

**Table 5: Employment Growth, Output Growth, Employment Elasticity and the ICT Intensity (%) in the ICT Using Services Sector during 2000-10**

| NIC 2004 | | | E | Y | EE | ICT/Non ICT (%) |
|---|---|---|---|---|---|---|
| 511 | Whole sale on a fee or contract basis | 2000-10 | 2.66 | 10.59 | 0.25 | 3.2 |
| | | 2000-05 | 2.56 | 15.30 | 0.17 | 1.61 |
| | | 2005-10 | 2.78 | 4.69 | 0.59 | 5.19 |
| 512 | Whole sale of agricultural raw material and live animals, food beverages and tobacco | 2000-10 | 6.29 | 11.59 | 0.54 | 5.41 |
| | | 2000-05 | 10.17 | 17.00 | 0.60 | 2.05 |
| | | 2005-10 | 1.43 | 4.83 | 0.30 | 9.6 |
| 513 | Wholesales of household goods | 2000-10 | 0.79 | 18.08 | 0.04 | 7.16 |
| | | 2000-05 | 0.25 | 18.53 | 0.01 | 12.49 |
| | | 2005-10 | 1.17 | 18.52 | 0.06 | 0.48 |
| 514 | Wholesale of non agricultural intermediate products, wastes and scraps | 2000-10 | 1.79 | 5.32 | 0.34 | 8.74 |
| | | 2000-05 | 2.25 | 4.65 | 0.48 | 4.06 |
| | | 2005-10 | 1.48 | 6.60 | 0.22 | 14.6 |
| 515 | Wholesales of machinery, equipment and supplies | 2000-10 | 5.43 | 9.21 | 0.59 | 10.1 |
| | | 2000-05 | 6.71 | 7.15 | 0.94 | 6.48 |
| | | 2005-10 | 5.34 | 11.80 | 0.45 | 14.62 |
| 519 | Other wholesales | 2000-10 | 3.15 | 4.17 | 0.76 | 5.64 |
| | | 2000-05 | 3.36 | 5.27 | 0.64 | 2.49 |
| | | 2005-10 | 6.64 | 8.79 | 0.76 | 9.58 |
| 521 | Retail sales in non specialized stores | 2000-10 | 14.46 | 22.99 | 0.63 | 28.2 |
| | | 2000-05 | 19.52 | 25.99 | 0.75 | 11.42 |
| | | 2005-10 | 8.12 | 19.25 | 0.42 | 49.18 |
| 526 | Repair of personal and household goods | 2000-10 | 4.80 | 7.14 | 0.67 | 14.03 |
| | | 2000-05 | 6.36 | 8.62 | 0.74 | 9.7 |
| | | 2005-10 | 4.10 | 6.54 | 0.63 | 19.43 |
| 621 | Scheduled Air Transport | 2000-10 | 0.06 | 7.20 | 0.01 | 3.69 |
| | | 2000-05 | 1.52 | 5.48 | 0.28 | 4.06 |
| | | 2005-10 | 2.03 | 9.35 | 0.22 | 3.23 |
| 603 | Transport via pipelines | 2000-10 | 4.10 | 19.77 | 0.21 | 0.93 |
| | | 2000-05 | 3.66 | 16.36 | 0.22 | 1.14 |
| | | 2005-10 | 4.66 | 23.02 | 0.20 | 0.67 |
| 711 | Renting of automobiles | 2000-10 | 1.85 | 13.76 | 0.13 | 0.55 |
| | | 2000-05 | 3.73 | 12.88 | 0.29 | 0.55 |
| | | 2005-10 | 4.82 | 15.31 | 0.31 | 0.54 |
| 731 | R&D on natural science and engineering | 2000-10 | 40.18 | 44.91 | 0.89 | 12.41 |
| | | 2000-05 | 56.10 | 67.57 | 0.83 | 8.21 |
| | | 2005-10 | 20.28 | 16.58 | 1.22 | 17.65 |
| 741 | Legal, accounting, book keeping and auditing activities; fiscal and management consulting | 2000-10 | 5.50 | 9.36 | 0.59 | 18.61 |
| | | 2000-05 | 11.12 | 9.31 | 1.19 | 9.16 |
| | | 2005-10 | -1.52 | 9.43 | -0.16 | 30.43 |
| 742 | Architectural and engineering activities and related technical consulting | 2000-10 | 9.68 | 20.24 | 0.48 | 24.82 |
| | | 2000-05 | 11.67 | 16.12 | 0.72 | 10.52 |
| | | 2005-10 | 10.19 | 24.40 | 0.42 | 42.69 |
| 743 | Advertising | 2000-10 | 3.55 | 10.26 | 0.35 | 26.71 |
| | | 2000-05 | 3.59 | 9.42 | 0.38 | 13.79 |
| | | 2005-10 | 3.48 | 12.06 | 0.29 | 42.86 |
| 749 | Business activities n.e.c. | 2000-10 | 18.67 | 20.21 | 0.92 | 14.43 |
| | | 2000-05 | 22.55 | 24.28 | 0.93 | 13.53 |
| | | 2005-10 | 14.83 | 15.12 | 0.98 | 15.54 |
| 803 | Business activities n.e.c. | 2000-10 | 5.74 | 12.00 | 0.48 | 7.17 |
| | | 2000-05 | 7.07 | 12.33 | 0.57 | 7.18 |
| | | 2005-10 | 6.84 | 11.27 | 0.61 | 7.15 |
| 851 | Human Health Activities | 2000-10 | 9.50 | 13.17 | 0.72 | 2.99 |
| | | 2000-05 | 10.99 | 13.49 | 0.81 | 1.5 |
| | | 2005-10 | 7.64 | 12.78 | 0.60 | 4.86 |
| 921 | Motion picture and video activities, radio and television activities | 2000-10 | 7.97 | 12.99 | 0.61 | 6.45 |
| | | 2000-05 | 7.48 | 14.21 | 0.53 | 5.2 |
| | | 2005-10 | 8.59 | 11.47 | 0.75 | 8.01 |
| 922 | News agency activities | 2000-10 | 9.24 | 16.30 | 0.57 | 6.78 |
| | | 2000-05 | 6.49 | 14.70 | 0.44 | 4.48 |
| | | 2005-10 | 12.68 | 18.30 | 0.69 | 9.65 |
| Aggregate | | 2000-10 | 3.58 | 12.85 | 0.28 | 5.15 |
| | | 2000-05 | 3.73 | 10.37 | 0.36 | 3.9 |
| | | 2005-10 | 3.29 | 12.69 | 0.26 | 6.72 |

Source: Own Computation using the Proves data source compiled by CMIE

**Table 6: Employment Growth, Output Growth, Employment Elasticity and the ICT Intensity (%) in the Non ICT Using Secondary Sector during 2000-10**

| NIC 2004 | | | E | Y | EE | ICT/Non ICT (%) |
|---|---|---|---|---|---|---|
| 10 | Mining and quarrying | 2000-10 | 1.29 | 12.27 | 0.11 | 0.01 |
| | | 2000-05 | 2.09 | 7.24 | 0.29 | 0 |
| | | 2005-10 | 0.3 | 18.56 | 0.02 | 0.03 |
| 152 | Dairy products | 2000-10 | 3.71 | 7.64 | 0.49 | 0.17 |
| | | 2000-05 | 4.63 | 10.25 | 0.45 | 0.06 |
| | | 2005-10 | 2.56 | 4.38 | 0.58 | 0.31 |
| 153 | Grain mill products, starch and starch products, prepared animal feeds | 2000-10 | 1.61 | 12.71 | 0.13 | 1.42 |
| | | 2000-05 | 1.42 | 15.85 | 0.09 | 0.77 |
| | | 2005-10 | 5.41 | 8.77 | 0.62 | 2.24 |
| 154 | Other food products | 2000-10 | 0.16 | 5.34 | 0.03 | 0.63 |
| | | 2000-05 | 2.37 | 5.57 | 0.43 | 0.27 |
| | | 2005-10 | 3.31 | 5.04 | 0.66 | 1.09 |
| 155 | Beverages | 2000-10 | 8.64 | 12.43 | 0.7 | 1.08 |
| | | 2000-05 | 8.56 | 13.81 | 0.62 | 0.35 |
| | | 2005-10 | 8.74 | 10.71 | 0.82 | 2 |
| 160 | Tobacco products | 2000-10 | 1.79 | 5.54 | 0.32 | 0.03 |
| | | 2000-05 | 0.42 | 4.65 | 0.09 | 0.03 |
| | | 2005-10 | 3.5 | 6.65 | 0.53 | 0.03 |
| 171 | Preparation and Spinning of textile fibers, textile weaving, made up textile ,expect apparel | 2000-10 | 2.31 | 4.96 | 0.47 | 0.45 |
| | | 2000-05 | 2.07 | 1.97 | 1.06 | 0.25 |
| | | 2005-10 | 2.61 | 8.71 | 0.3 | 0.7 |
| 172 | Other textiles | 2000-10 | 6.26 | 11.63 | 0.54 | 1.28 |
| | | 2000-05 | 2.93 | 10.45 | 0.28 | 0.66 |
| | | 2005-10 | 10.43 | 13.11 | 0.8 | 2.05 |
| 173 | Knitted and crocheted fabrics and articles | 2000-10 | 12.29 | 18.26 | 0.67 | 1.5 |
| | | 2000-05 | 10.51 | 19.45 | 0.54 | 0.84 |
| | | 2005-10 | 14.52 | 16.78 | 0.87 | 2.33 |
| 191 | Leather, Tanning and dressing of leather, luggage | 2000-10 | 3.1 | 9.18 | 0.33 | 5.68 |
| | | 2000-05 | 2.45 | 7.06 | 0.34 | 3.05 |
| | | 2005-10 | 3.43 | 12.82 | 0.22 | 8.97 |
| 192 | Manufacture of footwear | 2000-10 | 1.47 | 2.18 | 0.68 | 2.29 |
| | | 2000-05 | 0.98 | 0.91 | 1.07 | 1.48 |
| | | 2005-10 | 2.1 | 3.78 | 0.55 | 3.3 |
| 202 | veneer of sheet, plywood, particle board, fiber board, and other panel | 2000-10 | 3.47 | 8.78 | 0.39 | 0.7 |
| | | 2000-05 | 1.87 | 3.97 | 0.47 | 0.35 |
| | | 2005-10 | 10.15 | 14.8 | 0.69 | 1.13 |
| 241 | Basic chemicals | 2000-10 | -1.31 | 4.44 | -0.3 | 0.2 |
| | | 2000-05 | 0.61 | 5.17 | 0.12 | 0.08 |
| | | 2005-10 | -3.72 | 3.51 | -1.06 | 0.34 |
| 242 | Other chemical products | 2000-10 | 1.4 | 7.84 | 0.18 | 1.34 |
| | | 2000-05 | -1.47 | 6.18 | -0.24 | 0.61 |
| | | 2005-10 | 4.99 | 9.91 | 0.5 | 2.24 |
| 251 | Rubber products | 2000-10 | 0.53 | 8.22 | 0.06 | 0.18 |
| | | 2000-05 | 1.77 | 8.59 | 0.21 | 0.09 |
| | | 2005-10 | -1.02 | 7.76 | -0.13 | 0.3 |
| 252 | Plastic products | 2000-10 | 2.54 | 7.68 | 0.33 | 0.89 |
| | | 2000-05 | -0.15 | 4.7 | -0.03 | 0.32 |
| | | 2005-10 | 5.9 | 11.4 | 0.52 | 1.61 |
| 261 | Glass and glass products | 2000-10 | 6.48 | 10.53 | 0.61 | 0.78 |
| | | 2000-05 | 5.6 | 7.27 | 0.77 | 0.7 |
| | | 2005-10 | 7.09 | 14.61 | 0.48 | 0.89 |
| 269 | Non-metallic mineral products | 2000-10 | 0.75 | 8.27 | 0.09 | 0.2 |
| | | 2000-05 | 1.26 | 5.19 | 0.24 | 0.08 |
| | | 2005-10 | 0.1 | 12.12 | 0.01 | 0.35 |
| 271 | Basic irons and steels and ferro-alloys | 2000-10 | 3.53 | 12.94 | 0.27 | 0.01 |
| | | 2000-05 | 3.26 | 9.2 | 0.35 | 0.01 |
| | | 2005-10 | 4.73 | 16.38 | 0.29 | 0.01 |
| 272 | Basic precious and non-ferrous metals | 2000-10 | 1.87 | 9.08 | 0.21 | 0.19 |
| | | 2000-05 | 2.02 | 9.35 | 0.22 | 0.13 |
| | | 2005-10 | 1.7 | 8.75 | 0.19 | 0.26 |
| 273 | Casting and metals | 2000-10 | -4.85 | 6.88 | -0.7 | 1.35 |
| | | 2000-05 | -3.6 | 2.52 | -1.43 | 0.61 |
| | | 2005-10 | -6.41 | 12.34 | -0.52 | 2.27 |
| 281 | Structural products, tanks, reservoir and steam generators | 2000-10 | 1.03 | 13.46 | 0.08 | 1.63 |
| | | 2000-05 | 0.42 | 9.19 | 0.05 | 0.71 |
| | | 2005-10 | 1.78 | 18.79 | 0.09 | 2.78 |
| 314 | Acculators, primary cells and primary batteries | 2000-10 | 0.74 | 12.5 | 0.06 | 2.25 |
| | | 2000-05 | 1.75 | 5.67 | 0.31 | 1.25 |
| | | 2005-10 | -0.53 | 21.02 | -0.03 | 3.5 |

**Table 6: Employment Growth, Output Growth, Employment Elasticity and the ICT Intensity (%) in the Non ICT Using Secondary Sector during 2000-10 (Contd.)**

| | | | | | | |
|---|---|---|---|---|---|---|
| 341 | Motor vehicles | 2000-10 | 3.48 | 10.42 | 0.33 | 1.36 |
| | | 2000-05 | 4.75 | 10.63 | 0.45 | 0.61 |
| | | 2005-10 | 1.89 | 10.16 | 0.19 | 2.31 |
| 343 | Parts and accessories for motor vehicles and engines | 2000-10 | 2.33 | 12.5 | 0.19 | 1.2 |
| | | 2000-05 | 1.56 | 15.72 | 0.1 | 0.63 |
| | | 2005-10 | 3.29 | 8.46 | 0.39 | 1.9 |
| 352 | Railways and tramways locomotives and rolling stocks | 2000-10 | 2.23 | 9.74 | 0.23 | 1.27 |
| | | 2000-05 | 2.64 | 8.29 | 0.31 | 0.43 |
| | | 2005-10 | 1.8 | 10.04 | 0.18 | 2.32 |
| 402 | Gas (distribution of gaseous fuels through mains lines) | 2000-10 | 2.62 | 9.21 | 0.28 | 0.13 |
| | | 2000-05 | 3.09 | 3.04 | 1.02 | 0.08 |
| | | 2005-10 | 2.02 | 16.92 | 0.12 | 0.21 |
| 410 | Collection, purification and distribution of water | 2000-10 | 6.14 | 11.95 | 0.52 | 0.02 |
| | | 2000-05 | 7.63 | 10.9 | 0.72 | 0 |
| | | 2005-10 | 4.28 | 13.16 | 0.33 | 0.05 |
| 452 | Building of complete constructions or parts thereof; civil engineering | 2000-10 | 5.87 | 20.4 | 0.28 | 2.49 |
| | | 2000-05 | 6.85 | 19.82 | 0.34 | 2.02 |
| | | 2005-10 | 5.05 | 20.39 | 0.25 | 3.09 |
| 453 | Building installation | 2000-10 | 8.06 | 20.11 | 0.4 | 1.33 |
| | | 2000-05 | 7.97 | 15.11 | 0.53 | 0.4 |
| | | 2005-10 | 8.42 | 24.86 | 0.34 | 2.49 |
| Aggregate | | 2000-10 | 4.29 | 7.55 | 0.62 | 0.39 |
| | | 2000-05 | 4.14 | 6.88 | 0.6 | 0.18 |
| | | 2005-10 | 4.77 | 8.39 | 0.56 | 0.65 |

Source: Own Computation using the Proves data source compiled by CMIE

**Table 7: Employment Growth, Output Growth, Employment Elasticity and the ICT Intensity (%) in the Non ICT Using Services Sector during 2000-10**

| | | | E | Y | EE | ICT/Non ICT (%) |
|---|---|---|---|---|---|---|
| 551 | Hotels; camping sites and other provision of short-stay accommodation | 2000-10 | 3.63 | 14.69 | 0.25 | 1.06 |
| | | 2000-05 | 3.17 | 14.36 | 0.22 | 0.41 |
| | | 2005-10 | 4.21 | 15.1 | 0.28 | 1.88 |
| 552 | Restaurants, bars and canteens | 2000-10 | 11.84 | 15.72 | 0.75 | 3.2 |
| | | 2000-05 | 15.68 | 21.97 | 0.71 | 1.62 |
| | | 2005-10 | 7.03 | 7.9 | 0.89 | 5.18 |
| 601 | Transport via railways | 2000-10 | 10.3 | 40.27 | 0.26 | 0.15 |
| | | 2000-05 | 4.41 | 22.32 | 0.20 | 0.08 |
| | | 2005-10 | 17.65 | 62.71 | 0.28 | 0.24 |
| 602 | Other land transport | 2000-10 | 9.08 | 18.74 | 0.48 | 3.68 |
| | | 2000-05 | 14.21 | 30.99 | 0.46 | 1.87 |
| | | 2005-10 | 2.68 | 3.43 | 0.78 | 5.95 |
| 611 | Sea and coastal water transport | 2000-10 | 2.1 | 23.42 | 0.09 | 0.72 |
| | | 2000-05 | 2.98 | 30.14 | 0.10 | 0.74 |
| | | 2005-10 | 1 | 15.03 | 0.07 | 0.71 |
| 612 | Inland water transport | 2000-10 | 3.05 | 11.63 | 0.26 | 0.12 |
| | | 2000-05 | -1.6 | 12.63 | -0.13 | 0.09 |
| | | 2005-10 | 2.87 | 10.38 | 0.28 | 0.15 |
| 641 | Post and courier activities | 2000-10 | 3.4 | 4.45 | 0.76 | 1 |
| | | 2000-05 | 1.62 | 19.24 | 0.08 | 1.13 |
| | | 2005-10 | 5.62 | -14.03 | -0.40 | 0.84 |
| 701 | Real estate activities with own or leased property | 2000-10 | 16.05 | 43.28 | 0.37 | 0.06 |
| | | 2000-05 | -2.78 | 29.46 | -0.09 | 0.05 |
| | | 2005-10 | 39.59 | 60.57 | 0.65 | 0.08 |
| 711 | Renting of transport equipment | 2000-10 | 0.5 | 21.6 | 0.02 | 0.55 |
| | | 2000-05 | -13.54 | -6.23 | 2.17 | 0.55 |
| | | 2005-10 | 18.05 | 56.39 | 0.32 | 0.54 |
| 801 | Primary education | 2000-10 | 7.73 | 17.85 | 0.43 | 3.4 |
| | | 2000-05 | 6.58 | 17.74 | 0.37 | 0.62 |
| | | 2005-10 | 7.92 | 18.59 | 0.43 | 6.88 |
| Aggregated | | 2000-10 | 4.5 | 15.75 | 0.28 | 0.45 |
| | | 2000-05 | 4.43 | 15.08 | 0.29 | 0.24 |
| | | 2005-10 | 4.6 | 16.49 | 0.27 | 0.70 |

Source: Own Computation using the Proves data source compiled by CMIE

## References:

- Die Bold, J. 1952. Automation: The Advent of the Automated Factory, New York, Van Nostrand.
- Freeman, C. and Soete, L. 1987. Technical Change and Full Employment, Blackwell, Oxford.
- Friedman, A. and Cornford, D. 1989. Computer System Development: History, Organization and Implementation, Chi Chester, Wiley.
- Gill, C. 1985. Work, Employment and the New Technology, Polity Press, Cambridge.
- Gordon, R. J. 2000. Does the 'New Economy' Measure up to the Great Inventions of the Past? *Journal of Economic Perspectives*, vol. 14, No. 4, Fall.
- G.O.I. 2003. Statistical Abstract (India), New Delhi.
- Hamermesh, D. 1993. Labour Demand, Princeton University Press, Princeton.
- Jenkins, C. and Sherman, B. 1979.The Collapse of Work, London, Eyre Methuen.
- Kuznets, S. 1972. Innovations and Adjustment in Economic Growth, *Swedish Journal of Economics*, No. 74, pp. 431-51.
- Luc, Soete. 1987. Employment, Unemployment and Technical Change: A Review of Economic Debate. In Freeman, C. and Luc, Soete*, Technical Change and Full Employment* (ed.), Basil Blackwell Publication, London.
- Kumar, A. 2005. Factors Underlying Jobless Growth in India and the Need for a New Development Paradigm, Bhartiya Samajik Chintan, January-March, 3 (4), pp. 215-29.
- Maddison, A. 1991. Dynamic Forces in Capitalist Development: Along term Comparative View, Oxford University Press, Oxford.
- Matteucci, N and Sterlacchini, A. 2003. ICT and Employment Growth in Italian Industries. Universita Politecnica della Marche, Italy, November.
- Morawetz, D. 1974. Employment Implication of Industrialization in Developing Countries, *Economic Journal*, Vol. 84, No.335, pp. 491-542.
- NASSCOM, Various Years. *Annual Report*, New Delhi.
- NASSCOM-Crisil Report, 2007. New Delhi.
- Nelson, R. 1993. National System of Innovations: A Comparative Study, Oxford University Press, Oxford.

- Oliner, S. D. and Sichel, D. E. 2000. The Resurgence of Growth in the Late 1990s: Is Information Technology the Story? *Journal of Economic Perspectives* Vol. 14, No. 4, Fall.
- OECD, 1996: Oslo Manual, 2nd Edition, Paris OECD.
- OECD. 2010. ICT in India: Performance, Growth and Key Challenges, DSTI/ICCP/IE (2008)7 (Final).
- Pianta, M. 2000. The Employment Impact of Product and Process Innovations. In Vivarelli, M. and M. Pianta (ed.), *The Employment Impact of Innovation: Evidence and Policy*, Rutledge, London, pp. 77-95.
- Pissarides, C.A. and G. Vallanti. 2003. Productivity Growth and Employment: Theory and Panel Estimates, Centre for Economic Performance, London School of Economics, London.
- Rifkin, J. 1995. The End of Work, G. P. Putnam's Sons, New York.
- Rogoff Kenneth, 2012: World Economic Forum. https://www.weforum.org/agenda/2012/10/king-ludd-is-still-dead
- Schmidt, H. 1983. Technological Change, Employment and Occupational qualifications', CEDEFOP, Vocational Training, Berlin.
- Soete, L. 1985. *Technological trend and Employment*, Vol No.5, Electronics and Communications.
- Stiroh J. 2002. Information Technology and the U.S. Productivity Revival: What Do the Industry Data Say? *The American Economic Review*, Vol. 92, No. 5, pp. 1559-1576.
- Taylor, A. et. al. 1985. The Impact of New Technology on Local Employment, Gower Publishing Company, Brookfield, USA.
- Vivarelli, M. et. al. 1996. Innovation and Employment in the Italian Manufacturing Industries, *Research Policy*, 25, pp. 1013-1026.
- Vivarelli, M. and M. Pianta. 2000. The Employment Impact of Innovation: Evidence and Policy, Routledge, London/ New York.
- Vivarelli, M. 2011. Innovation, Employment and Skill in Advanced and Developing Countries, Inter American Development Bank, Technical Notes, No. IDB-TN-351.

- Whitley, J.D. and Wilson, R.A. 1981. Quantifying the Employment Impact of Micro-electronics', Institute for Employment Research, Warwick, Discussion Paper No. 15, December.